\documentclass[prb,preprint,showpacs,float]{revtex4-1}
\usepackage{graphicx}
\usepackage{amssymb}
\usepackage{amsmath}
\usepackage{xspace}
\usepackage{url}
\usepackage{subfigure}

\begin{document} 
\title{Electron flow in circular graphene quantum dots}

\author{C. Schulz, R. L. Heinisch, and 
H. Fehske}
\affiliation{Institut f{\"ur} Physik,
             Ernst-Moritz-Arndt-Universit{\"a}t Greifswald,
             17487 Greifswald,
             Germany}

\date{\today}
\begin{abstract}
We study the electron propagation in a circular electrostatically defined quantum dot in graphene. Solving the scattering problem for a plane Dirac electron wave we identify different scattering regimes depending on the radius and potential of the dot as well as the electron energy.  Thereby, our particular focus lies on resonant scattering and the temporary electron trapping in the dot.   
 
\end{abstract}
\maketitle

\section{introduction}

One of the most intriguing properties of graphene is the quasi-relativistic propagation of electrons.
The intersection of energy bands at the edge of the Brillouin zone leads to a gapless conical energy spectrum in two inequivalent valleys. Low-energy quasiparticles are described by a massless Dirac equation\cite{Dirac28} and have a pseudospin which gives the contribution of the two sublattices of the graphene honeycomb lattice to their make-up. Chirality---the projection of the pseudospin on the direction of motion---is  responsible for the conservation of pseudospin  and the absence of backscattering  for potentials diagonal in sublattice space.\cite{ANS98} This particularity of the band structure allows perfect transmission for normal incidence at a n-p junction, termed Klein tunnelling. \cite{Klein28,KNG06} 

As Klein tunnelling prevents the electrostatic confinement of electrons, proposals for circuitry have---in view of the linear energy dispersion---been modelled on optical analogues.\cite{CFA07,PSY08,WLL11} Most notably n-p junctions could serve as lenses refocussing diverging electron rays.\cite{CFA07}
Incidentally, the experimental verification of Klein tunnelling was obtained in a Fabry-Perot interferometer composed of two n-p junctions.\cite{YK09}

In this work we turn to the electron flow in a circular electrostatically defined quantum dot in graphene. For large dots---adequately described by ray optics---diffraction at the boundary results in two caustics inside the dot.\cite{CPP07,AU13} For a small dot, however, wave optical features emerge.  In this regime resonances in the conductance \cite{BTB09,PAS11} and the scattering cross section \cite{HBF13a,PHF13} indicate quasi-bound states at the dot. Indeed, electrons can be confined in a circular dot surrounded by unbiased graphene as the classical electron dynamics in the dot is integrable and the corresponding Dirac equation is separable.  \cite{HA08,HA09,BTB09} 

In the following, we consider the scattering of a plane Dirac electron wave on a circular potential step. We study different scattering regimes, with a particular focus on quasi-bound states occurring for resonant scattering.

\section{theory}

For a dot potential that is smooth on the scale of the lattice constant but sharp on the scale of the de Broglie wavelength the low-energy electron dynamics in graphene is described by the single-valley Dirac-Hamiltonian 
\begin{equation}
H=-\text{i}\nabla\mathbf{\sigma}+V\theta (R-r),
\end{equation}
 where $R$ is the radius, \(V\) the applied bias of the gated region, and \(\mathbf{\sigma}=(\sigma_x,\sigma_y)\) are Pauli matrices. We use units such that \(\hbar=1\) and the Fermi  velocity \(v_\mathrm{F}=1\). 
  
In polar coordinates the Hamiltonian of unbiased graphene ($V=0$) takes the form 
\begin{equation}
H=\left(\begin{array}{cc}
0 & \textrm{e}^{-\textrm{i}\varphi}\left(-\textrm{i}\frac{\partial}{\partial r}-\frac{1}{r}\frac{\partial}{\partial\varphi}\right)\\
\textrm{e}^{\textrm{i}\varphi}\left(-\textrm{i}\frac{\partial}{\partial r}+\frac{1}{r}\frac{\partial}{\partial\varphi}\right) & 0\end{array}\right).
\end{equation}
As a consequence of the spinor structure of the Hamiltonian the angular orbital momentum $L=x \frac{1}{\text{i}} \frac{\partial}{\partial y}-y \frac{1}{\text{i}} \frac{\partial}{\partial x}$ does not commute with the Hamiltonian, that is $[L,H]=\text{i}\left(\sigma_{x} \frac{1}{\text{i}} \frac{\partial}{\partial y}-\sigma_{y}\frac{1}{\text{i}} \frac{\partial}{\partial x}\right)$. However, the total angular momentum including orbital angular momentum as well as pseudo-spin, $J=L+\sigma_z/2$, satisfies $[J,H]=0$. 

To construct the eigenfunctions of the free Dirac equation in polar coordinates, $H \psi(r,\varphi)=E\psi(r,\varphi)$, we need the eigenfunctions of $J$. The eigenfunctions of $J$ to the eigenvalue $m+1/2$ are given by
\begin{equation}
\xi_m^+=\frac{e^{\text{i}m\varphi}}{\sqrt{2\pi}}\left(\begin{array}{c}
1\\
0\end{array}\right), \quad \xi_{m+1}^-=\frac{e^{\text{i}(m+1)\varphi}}{\sqrt{2\pi}}\left(\begin{array}{c}
0\\
1\end{array}\right).
\end{equation}
In the next step we obtain the eigenfunction to $J$ which is also an eigenfunction to $H$. For this, we make the ansatz
\begin{equation}\label{126}
\psi_{m}\left(r,\varphi\right)=f_{m}^{+}\left(r\right) \xi_{m}^{+}\left(\varphi\right)+f_{m+1}^{-}\left(r\right) \xi_{m+1}^{-}\left(\varphi\right).
\end{equation}
Using the relation
\begin{equation}\label{128}
Hf_{m}^{\pm}\left(r\right)\xi_{m}^{\pm}\left(\varphi\right)=\left(-\textrm{i}\frac{\partial}{\partial r}f_{m}^{\pm}\left(r\right)\pm\textrm{i}\frac{m}{r}f_{m}^{\pm}\left(r\right)\right)\xi_{m\pm1}^{\mp}\,,
\end{equation}
we find the set of equations
\begin{equation}\label{129}
-\textrm{i}\frac{\partial}{\partial r}f_{m}^{+}\left(r\right)+\textrm{i}\frac{m}{r}f_{m}^{+}\left(r\right)=Ef_{m+1}^{-}\left(r\right),
\end{equation}
\begin{equation}\label{130}
-\textrm{i}\frac{\partial}{\partial r}f_{m+1}^{-}\left(r\right)-\textrm{i}\frac{m+1}{r}f_{m+1}^{-}\left(r\right)=Ef_{m}^{+}\left(r\right),
\end{equation}
which determine the radial wave functions $f_{m}^{+}\left(r\right)$ and $f_{m+1}^{-}\left(r\right)$. Thereby, we have separated the Dirac equation in radial and angular parts. Substituting $z=Er,$ $f_{m+1}^{-}\left(z\right)=Z_{m+1}\left(z\right)$ and $\text{i}f_{m}^{+}\left(z\right)=Z_{m}\left(z\right)$, the above equations lead to the recurrence relations of the Bessel functions $Z_m$. The eigenfunction of $H$ in polar coordinates are thus given by
\begin{equation}
\psi_m(r,\varphi)=-\textrm{i}Z_{m}\left(kr	\right)\frac{1}{\sqrt{2\pi}}\textrm{e}^{\textrm{i}m\varphi}\left(\begin{array}{c}
1\\
0\end{array}\right)+Z_{m+1}\left(kr\right)\frac{1}{\sqrt{2\pi}}\textrm{e}^{\textrm{i}\left(m+1\right)\varphi}\left(\begin{array}{c}
0\\
1\end{array}\right),
\end{equation}
where the wave number $k=E$.

\begin{figure}
\begin{minipage}{0.49\linewidth}
\includegraphics[width=1\linewidth]{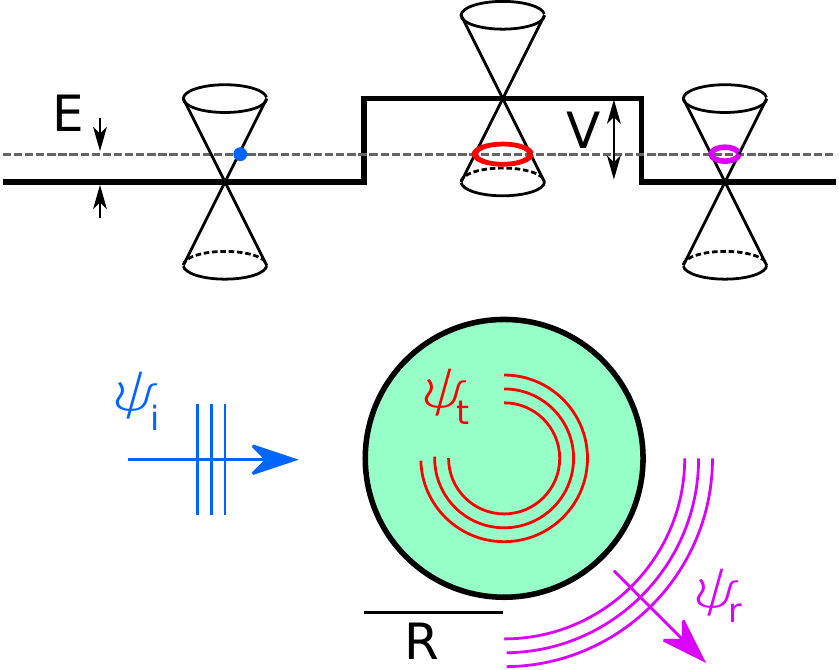}
\end{minipage}
\begin{minipage}{0.49\linewidth}
\caption{Sketch of the scattering geometry and the band structure for the scattering of a low-energetic electron at a circular dot in graphene. The dot is characterised by its radius $R$ and the applied bias $V$. The incident plane wave with energy $E>0$ (blue) corresponds to a state in the conduction band (upper cone). The reflected wave (purple) also lies in the conduction band while the transmitted wave inside the dot (red) occupies the valence band (lower cone).}\label{figure1}

\end{minipage}

\end{figure}

We now turn to the scattering of a plane wave on a circular dot. The scattering geometry is illustrated in Fig. \ref{figure1}. The wave function of the incident electron, assumed to propagate in the $x$-direction,  can be expanded according to
\begin{equation}\label{136}
\psi_{\textrm{i}}=\frac{1}{\sqrt{2}}\textrm{e}^{\textrm{i}kx}\left(\begin{array}{c}
1\\
1\end{array}\right)=\frac{1}{\sqrt{2}}\textrm{e}^{\textrm{i}kr \cos\varphi}\left(\begin{array}{c}
1\\
1\end{array}\right)=\frac{1}{\sqrt{2}}\sum_{m=-\infty}^{\infty}\textrm{i}^{m}J_{m}\left(kr\right)\textrm{e}^{\textrm{i}m\varphi}\left(\begin{array}{c}
1\\
1\end{array}\right)
\end{equation}
that is
\begin{equation}\label{137}
\psi_{\textrm{i}} =\sum_{m=-\infty}^{\infty}\sqrt{\pi}\textrm{i}^{m+1}\left[-\textrm{i}J_{m}\left(kr	\right)\frac{1}{\sqrt{2\pi}}\textrm{e}^{\textrm{i}m\varphi}\left(\begin{array}{c}
1\\
0\end{array}\right)+J_{m+1}\left(kr\right)\frac{1}{\sqrt{2\pi}}\textrm{e}^{\textrm{i}\left(m+1\right)\varphi}\left(\begin{array}{c}
0\\
1\end{array}\right)\right].
\end{equation}
This wave function is then matched with the reflected (scattered) and transmitted waves which are expanded in eigenfunctions in polar coordinates as well. The reflected wave is an outgoing wave. From the asymptotic behaviour of the Hankel function of the first kind $H_{m}^{\left(1\right)}\left(kr\right)\underset{_{kr\gg1}}{\sim}\sqrt{\frac{2}{\pi kr}}\textrm{e}^{\textrm{i}\left(kr-m\frac{\pi}{2}-\frac{\pi}{4}\right)}$ follows that for the reflected wave $Z_m(kr)=H_m^{(1)}(kr)$.
Introducing the scattering coefficients $a_m$, the reflected wave reads
\begin{equation}\label{139}
\psi_{\textrm{r}}=\sum_{m=-\infty}^{\infty}\sqrt{\pi}\textrm{i}^{m+1}a_{m}\left[-\textrm{i}H_{m}^{\left(1\right)}\left(kr\right)\frac{1}{\sqrt{2\pi}}\textrm{e}^{\textrm{i}m\varphi}\left(\begin{array}{c}
1\\
0\end{array}\right)+H_{m+1}^{\left(1\right)}\left(kr\right)\frac{1}{\sqrt{2\pi}}\textrm{e}^{\textrm{i}\left(m+1\right)\varphi}\left(\begin{array}{c}
0\\
1\end{array}\right)\right].
\end{equation}

The transmitted wave inside the dot has to be bounded at the origin. Only Bessel's functions $J_m$ satisfy this requirement. Thus the transmitted wave is given by
\begin{equation}\label{140}
\psi_{\textrm{t}}=\sum_{m=-\infty}^{\infty}\sqrt{\pi}\textrm{i}^{m+1}b_{m}\left[-\textrm{i}J_{m}\left(qr\right)\frac{1}{\sqrt{2\pi}}\textrm{e}^{\textrm{i}m\varphi}\left(\begin{array}{c}
1\\
0\end{array}\right)+\alpha J_{m+1}\left(qr\right)\frac{1}{\sqrt{2\pi}}\textrm{e}^{\textrm{i}\left(m+1\right)\varphi}\left(\begin{array}{c}
0\\
1\end{array}\right)\right],
\end{equation}
where the  $b_m$ denote the transmission coefficients and $\alpha=\textrm{sgn}\left(E-V\right)$ as well as $q=\alpha\left(E-V\right)$. 

Requiring continuity of the wave functions at the boundary of the dot, that is $\psi_{\textrm{i}}\left(r=R\right)+\psi_{\text{r}}\left(r=R\right)=\psi_{\textrm{t}}\left(r=R\right)$, gives the conditions
\begin{equation}\label{142}
J_{m}\left(kR\right)+a_{m}H_{m}^{\left(1\right)}\left(kR\right)=b_{m}J_{m}\left(qR\right),
\end{equation}
\begin{equation}\label{143}
J_{m+1}\left(kR\right)+a_{m}H_{m+1}^{\left(1\right)}\left(kR\right)=\alpha b_{m}J_{m+1}\left(qR\right).
\end{equation}
Solving these equations gives the scattering coefficients
\begin{equation}\label{144}
a_{m}=\frac{-J_{m}\left(N\rho\right)J_{m+1}\left(\rho\right)+\alpha J_{m}\left(\rho\right)J_{m+1}\left(N\rho\right)}{J_{m}\left(N\rho\right)H_{m+1}^{\left(1\right)}\left(\rho\right)-\alpha J_{m+1}\left(N\rho\right)H_{m}^{\left(1\right)}\left(\rho\right)}\,,
\end{equation}
\begin{equation}\label{145}
b_{m}=\frac{J_{m}\left(\rho\right)H_{m+1}^{\left(1\right)}\left(\rho\right)-J_{m+1}\left(\rho\right)H_{m}^{\left(1\right)}\left(\rho\right)}{J_{m}\left(N\rho\right)H_{m+1}^{\left(1\right)}\left(\rho\right)-\alpha J_{m+1}\left(N\rho\right)H_{m}^{\left(1\right)}\left(\rho\right)}\,,
\end{equation}
where we have introduced the size parameter $\rho=kR$ and the modulus of the refractive index $N=\alpha \left(E-V\right)/E$.
Using $J_{-m}=\left(-1\right)^{m}J_{m}$ and $ H_{-m}^{\left(1\right)}=\left(-1\right)^{m}H_{m}^{\left(1\right)}$ we find $a_{-m}=a_{m-1}$ and $b_{-m}=\alpha b_{m-1}$. The expressions for the reflected and transmitted wave can thus be rewritten as
\begin{equation}\label{151}
\psi_{\textrm{r}}=\frac{1}{\sqrt{2}}\sum_{m=0}^{\infty}\textrm{i}^{m+1}a_{m}\left[-\textrm{i}H_{m}^{\left(1\right)}\left(kr\right)\left(\begin{array}{c}
\textrm{e}^{\textrm{i}m\varphi}\\
\textrm{e}^{-\textrm{i}m\varphi}\end{array}\right)+H_{m+1}^{\left(1\right)}\left(kr\right)\left(\begin{array}{c}
\textrm{e}^{-\textrm{i}\left(m+1\right)\varphi}\\
\textrm{e}^{\textrm{i}\left(m+1\right)\varphi}\end{array}\right)\right]\,,
\end{equation}
as well as
\begin{equation}\label{151x}
\psi_{\textrm{t}}=\frac{1}{\sqrt{2}}\sum_{m=0}^{\infty}\textrm{i}^{m+1}b_{m}\left[-\textrm{i}J_{m}\left(qr\right)\left(\begin{array}{c}
\textrm{e}^{\textrm{i}m\varphi}\\
\textrm{e}^{-\textrm{i}m\varphi}\end{array}\right)+\alpha J_{m+1}\left(qr\right)\left(\begin{array}{c}
\textrm{e}^{-\textrm{i}\left(m+1\right)\varphi}\\
\textrm{e}^{\textrm{i}\left(m+1\right)\varphi}\end{array}\right)\right]\,.
\end{equation}

The electron density is given by $n=\psi^\dagger \psi$ and the current by $\mathbf{j}=\psi^\dagger \sigma \psi$ where $\psi=\psi_\text{i}+\psi_\text{r}$  outside and $\psi=\psi_\text{t}$ inside the dot.
The radial component of the current is given by
\begin{equation}\label{153}
j_{r}=\psi^\dagger \left[ \sigma_{x}\textrm{cos}\varphi+\sigma_{y}\textrm{sin}\varphi \right] \psi =\psi^\dagger \left[ \left(\begin{array}{cc}
0 & 1\\
1 & 0\end{array}\right)\textrm{cos}\varphi+\left(\begin{array}{cc}
0 & -\textrm{i}\\
\textrm{i} & 0\end{array}\right)\textrm{sin}\varphi \right] \psi.
\end{equation}
Thus for the reflected wave the radial current takes the form
\begin{equation}\label{154}
 \begin{split}
j_{r}^{\textrm{r}}\left(\varphi\right)
									 & =\frac{1}{2} \sum_{m=0}^{\infty}\left(-\textrm{i}\right)^{m+1}a_{m}^{*}\left[\textrm{i}H_{m}^{\left(1\right)*}\left(kr\right)\left(\begin{array}{c}
\textrm{e}^{-\textrm{i}m\varphi}\\
\textrm{e}^{\textrm{i}m\varphi}\end{array}\right)+H_{m+1}^{\left(1\right)*}\left(kr\right)\left(\begin{array}{c}
\textrm{e}^{\textrm{i}\left(m+1\right)\varphi}\\
\textrm{e}^{-\textrm{i}\left(m+1\right)\varphi}\end{array}\right)\right] \\
									 & \times\sum_{n=0}^{\infty}\textrm{i}^{n+1}a_{n}\left[-\textrm{i}H_{n}^{\left(1\right)}\left(kr\right)\left(\begin{array}{c}
\textrm{e}^{-\textrm{i}\left(n+1\right)\varphi}\\
\textrm{e}^{\textrm{i}\left(n+1\right)\varphi}\end{array}\right)+H_{n+1}^{\left(1\right)}\left(kr\right)\left(\begin{array}{c}
\textrm{e}^{\textrm{i}n\varphi}\\
\textrm{e}^{-\textrm{i}n\varphi}\end{array}\right)\right].
\end{split}
\end{equation}
Simplifying this leads to
\begin{equation}\label{155}
\begin{split}
j_{r}^{\textrm{r}}(\varphi)=\sum_{m,n=0}^{\infty}a_{m}^{*}a_{n}\textrm{i}^{n-m}&\left[\left(H_{m}^{\left(1\right)*}\left(kr\right)H_{n}^{\left(1\right)}\left(kr\right)+H_{m+1}^{\left(1\right)*}\left(kr\right)H_{n+1}^{\left(1\right)}\left(kr\right)\right)\textrm{cos}\left(\left(m+n+1\right)\varphi\right)\right. \\
									 &  +\left.\textrm{i}\left(H_{m}^{\left(1\right)*}\left(kr\right)H_{n+1}^{\left(1\right)}\left(kr\right)-H_{m+1}^{\left(1\right)*}\left(kr\right)H_{n}^{\left(1\right)}\left(kr\right)\right)\textrm{cos}\left(\left(m-n\right)\varphi\right)\right].
\end{split}
\end{equation}
In the far field $j_r^\text{r}(\varphi)$ gives the angular scattering characteristic. In this limit ($kr\rightarrow \infty$) we can evaluate the above expression using the asymptotic behaviour of the Hankel functions. We obtain
\begin{equation}\label{156}
 \begin{split}
j_{r}^{\textrm{r}}\left(\varphi\right)& =\frac{4}{\pi kr}\sum_{m=0}^{\infty}\left|a_{m}\right|^{2}\left[\textrm{cos}\left(\left(2m+1\right)\varphi\right)+1\right] \\
							 & + \frac{8}{\pi kr}\sum_{n<m}\textrm{Re}\left(a_{n}a_{m}^{*}\right)\left[\textrm{cos}\left(\left(m+n+1\right)									\varphi\right)+\textrm{cos}\left(\left(m-n\right)\varphi\right)\right].
\end{split}
\end{equation}

The scattering cross section $\sigma= I_r^\text{r} / (I^\text{i}/A_u)$ is given by the total reflected flux through a concentric circle $I_r^\text{r}$ divided by the incident flux per unit area $I^\text{i}/A_u$. As a consequence of continuity of the current $I_r^\text{r}$ can be calculated using the far field radial reflected current. This gives 
\begin{equation}\label{158}
I_{r}^{\textrm{r}}=\int_{0}^{2\pi}j_{r}^{\textrm{r}}r\textrm{d}\varphi=\frac{8}{k}\sum_{m=0}^{\infty}\left|a_{m}\right|^{2}.
\end{equation}
For the incident wave  $\psi_\text{i}=\frac{1}{\sqrt{2}}e^{ikx} \binom{1}{1}$  we obtain $I^\text{i}/A_u=1$.
To compare  scattering on dots of different size we use the scattering efficiency. This is the scattering cross section divided by the geometric cross section. It is given by
\begin{equation}\label{160}
Q=\frac{\sigma}{2R}=\frac{4}{\rho}\sum_{m=0}^{\infty}\left|a_{m}\right|^{2}.
\end{equation}

\section{results}

The complex mathematical expressions of the scattering coefficients encompass a variety of scattering phenomena depending on the electron energy $E$, the dot radius $R$ and the dot potential $V$. 

\begin{figure}
\begin{minipage}{0.49\linewidth}
\includegraphics[width=1\linewidth]{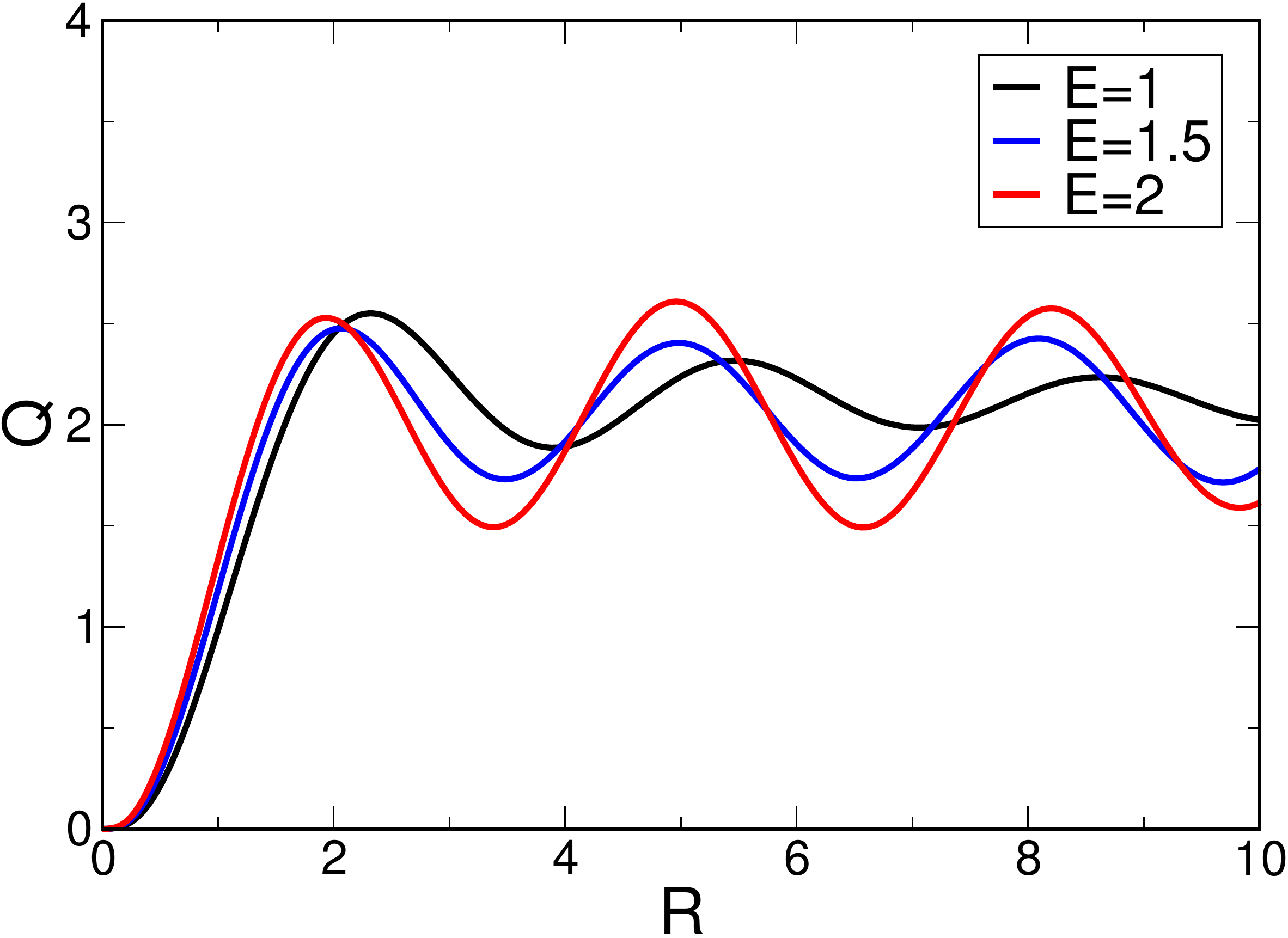}
\end{minipage}
\begin{minipage}{0.49\linewidth}
\includegraphics[width=1\linewidth]{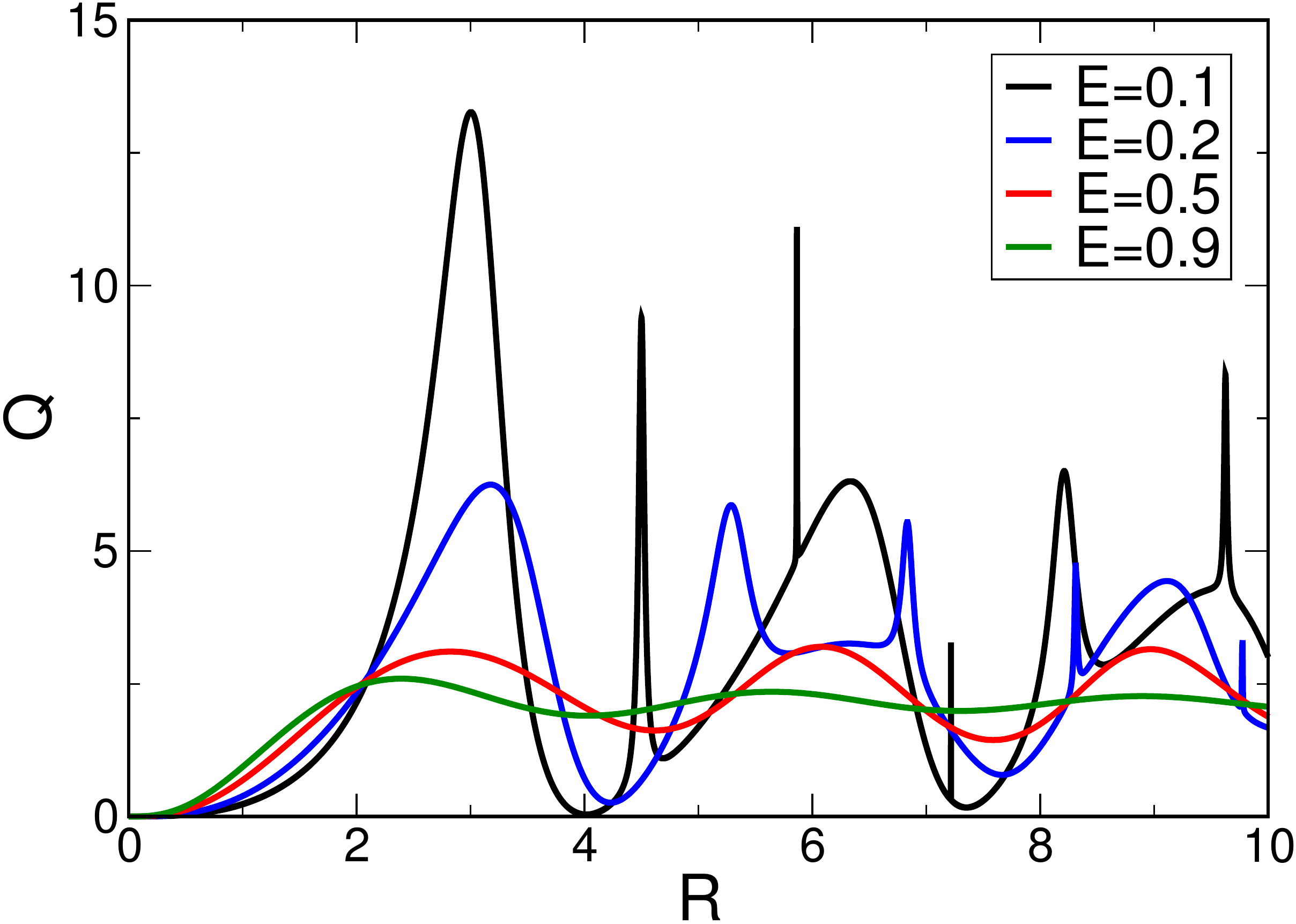}
\end{minipage}
\caption{Scattering efficiency $Q$ as a function of the radius of the dot $R$ for different energies of the incident electron $E$. The potential of the dot is $V=1$. The left panel gives results for $E \geq V$ (n-n junction), the right panel for $E<V$ (n-p junction).}
\label{figure2}
\end{figure}

Let us begin by analysing the scattering efficiency $Q$ as a function of $R$ for given $E$ and $V$. On the left panel of Fig. \ref{figure2} we show $Q(R)$ for over-the-barrier scattering, that is a n-n junction. The electron wave occupies outside and inside the dot states in the valence band. The scattering efficiency shows an oscillatory behaviour. For the case of threshold incidence, that is $E \rightarrow V$ the oscillations are damped. In this case the electronic state inside the dot is close to the Dirac point and the corresponding wavelength inside the dot diverges. The electron can thus scarcely penetrate into the dot. A Schr\"odinger electron with quadratic energy dispersion would qualitatively show a similar behaviour as extended states are present inside and outside the dot as well.

For $E<V$, that is under-the-barrier scattering (n-p junction), the particularity of the graphene energy dispersion becomes apparent. For a Schr\"odinger electron only evanescent waves exist inside the dot for $E<V$. As a consequence the scattering efficiency shows no or only strongly damped oscillatory behaviour. A Dirac electron, on the contrary, occupies inside the dot a state in the valence band. The right panel of Fig. \ref{figure2} shows the oscillatory behaviour of $Q(R)$ for $E<V$. For larger $E$ these oscillations are relatively smooth while for small $E$ their amplitude increases significantly and sharp peaks emerge. 

\begin{figure}
\begin{minipage}{0.49\linewidth}
\includegraphics[width=1\linewidth]{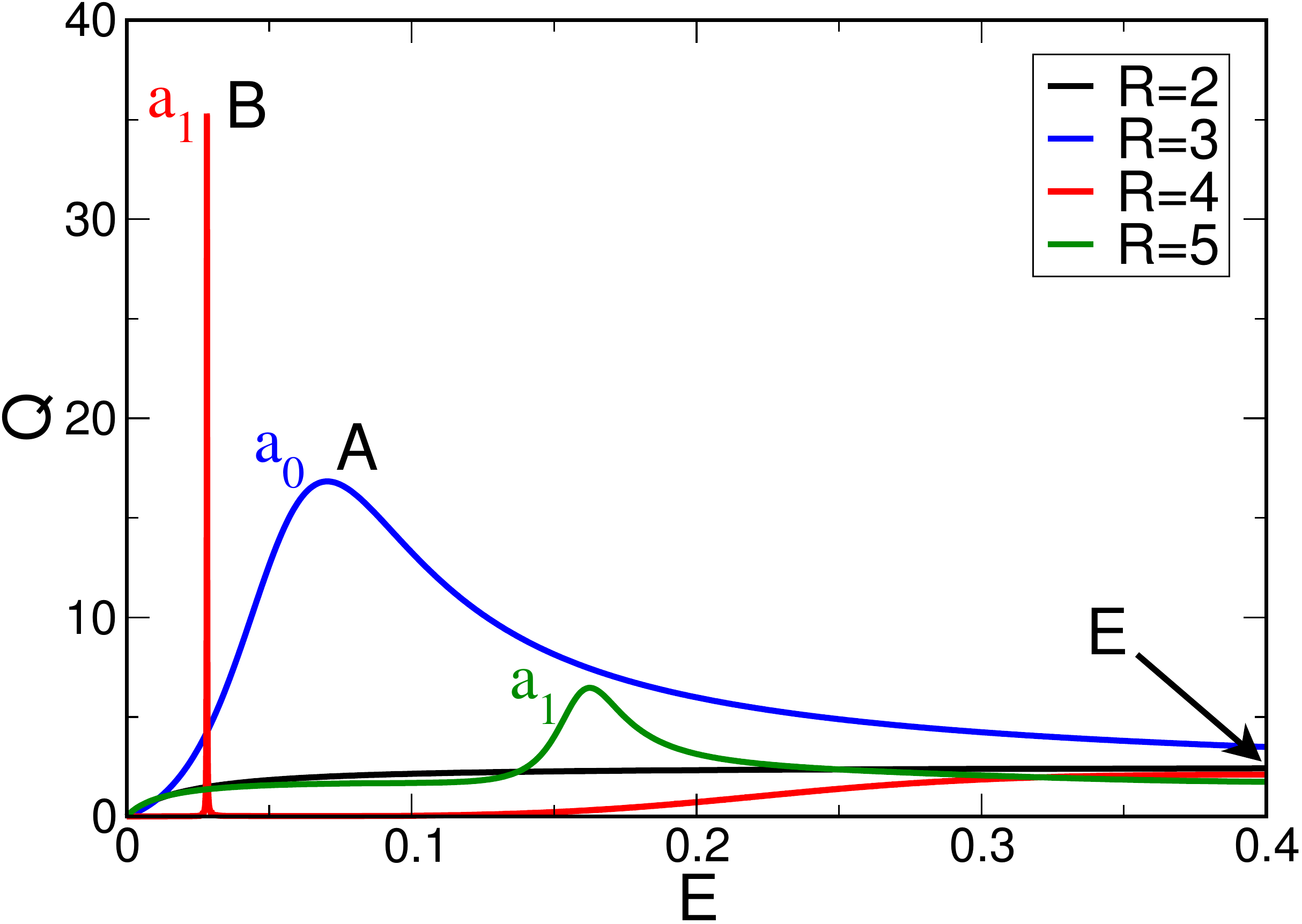}
\end{minipage}
\begin{minipage}{0.49\linewidth}
\includegraphics[width=1\linewidth]{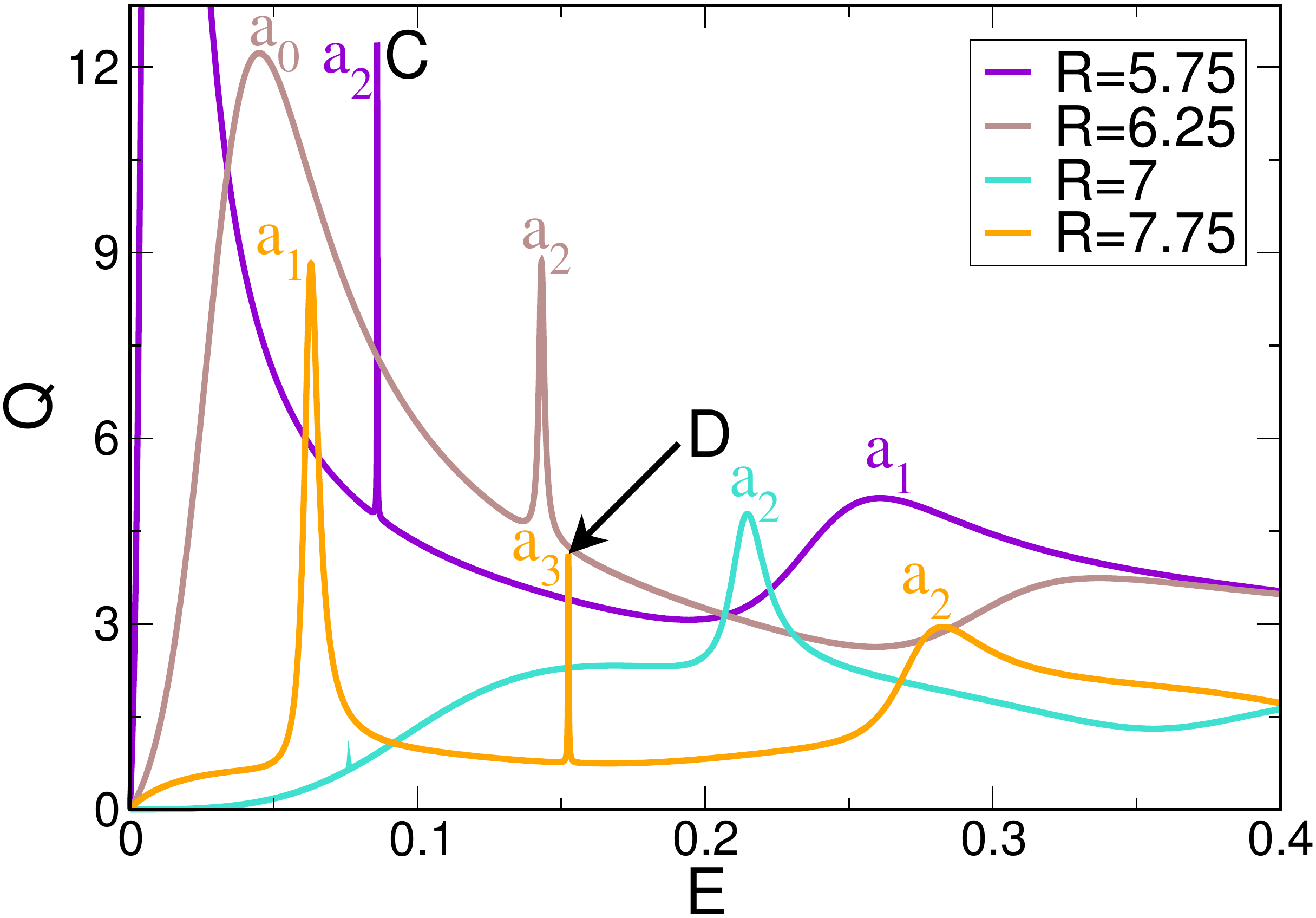}
\end{minipage}
\caption{Scattering efficiency $Q$ as a function of the energy $E$ of the incident electron for a dot with different radii $R$ and applied potential $V=1$. The scattering efficiency displays several resonances. The corresponding normal modes are indicated. Markers A-E denote particular cases which will be analysed further in the following.}
\label{figure3}
\end{figure}

\begin{figure}
\begin{minipage}{0.32\linewidth}
\includegraphics[width=1\linewidth]{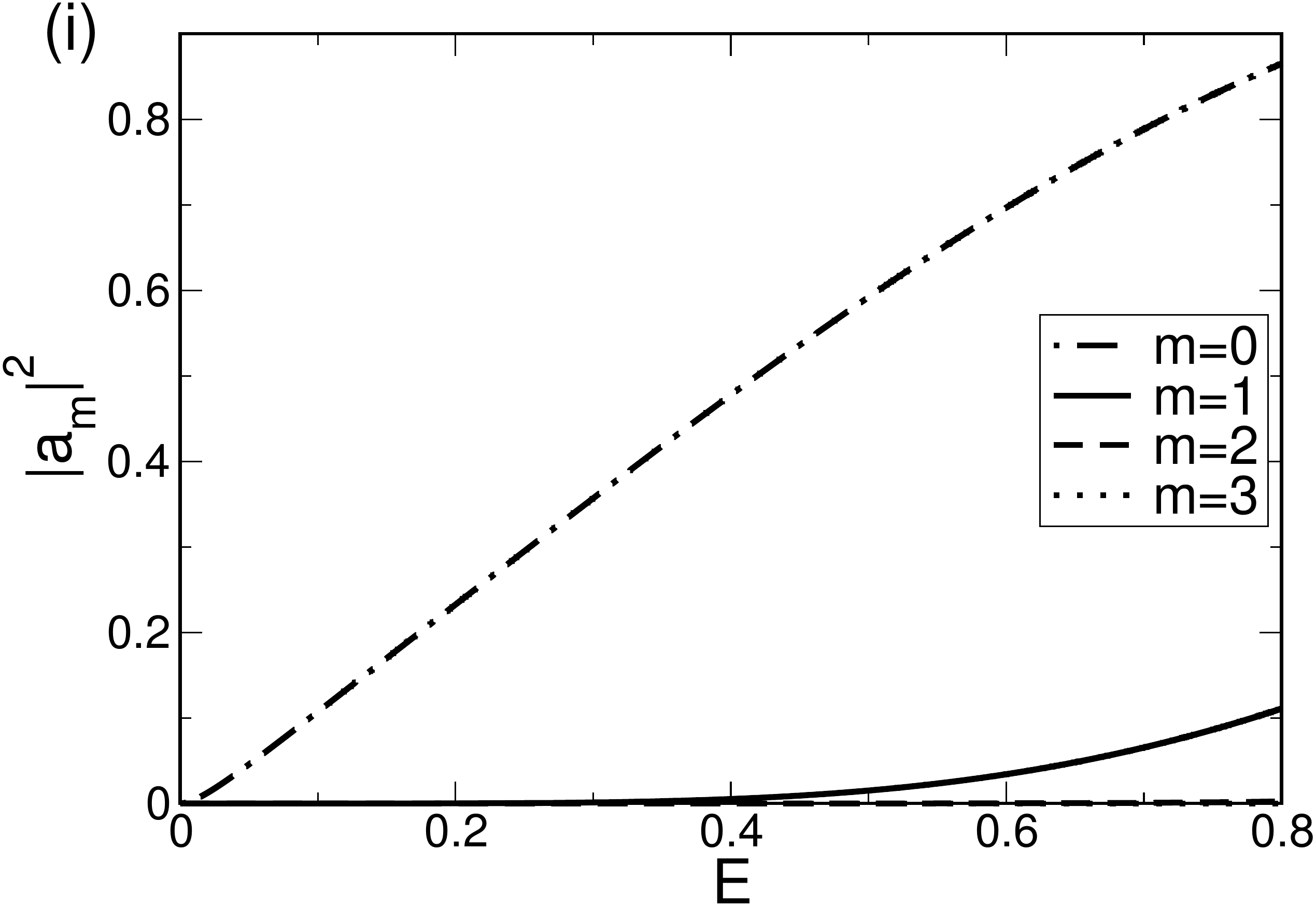}
\end{minipage}
\begin{minipage}{0.32\linewidth}
\includegraphics[width=1\linewidth]{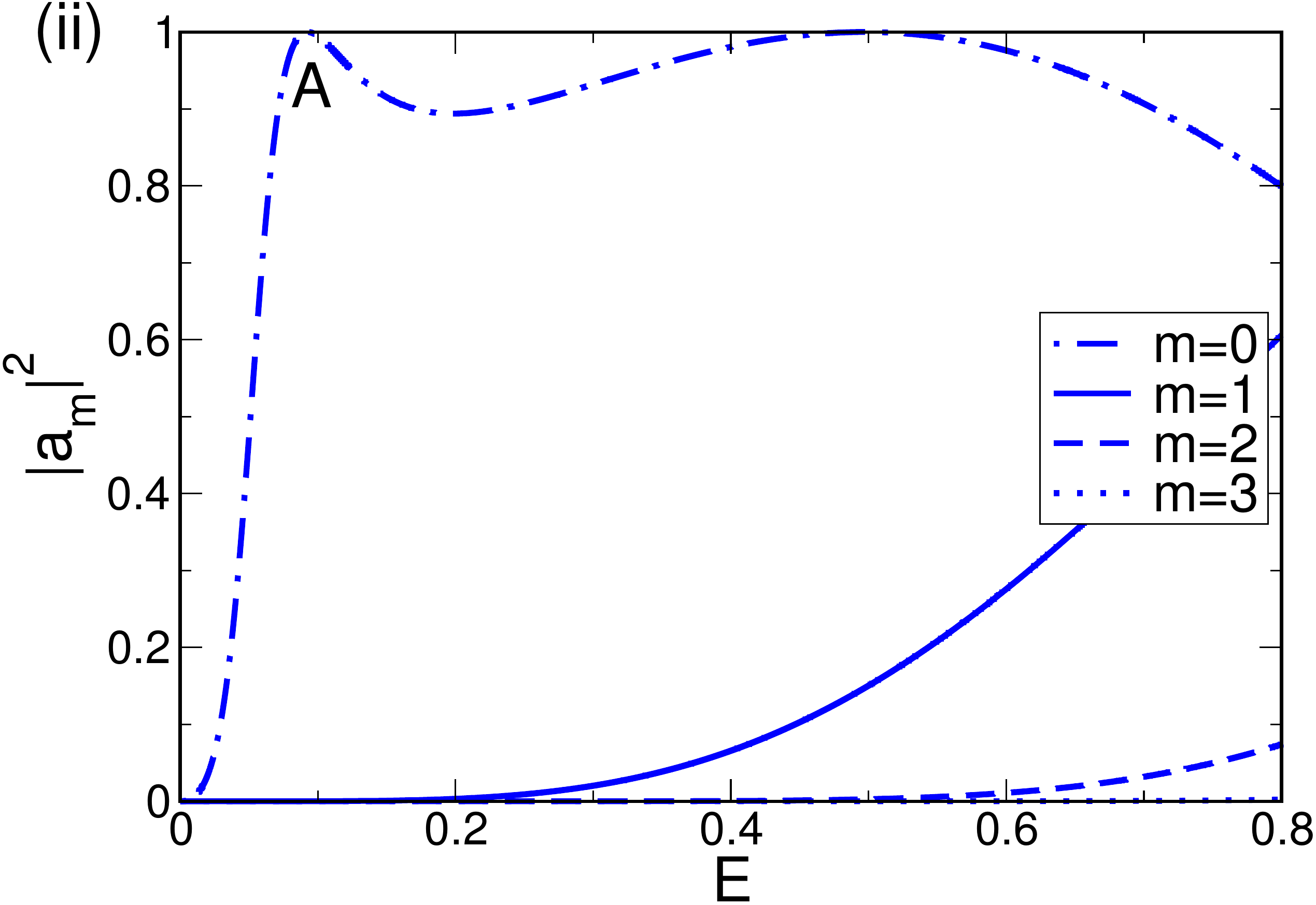}
\end{minipage}
\begin{minipage}{0.32\linewidth}
\includegraphics[width=1\linewidth]{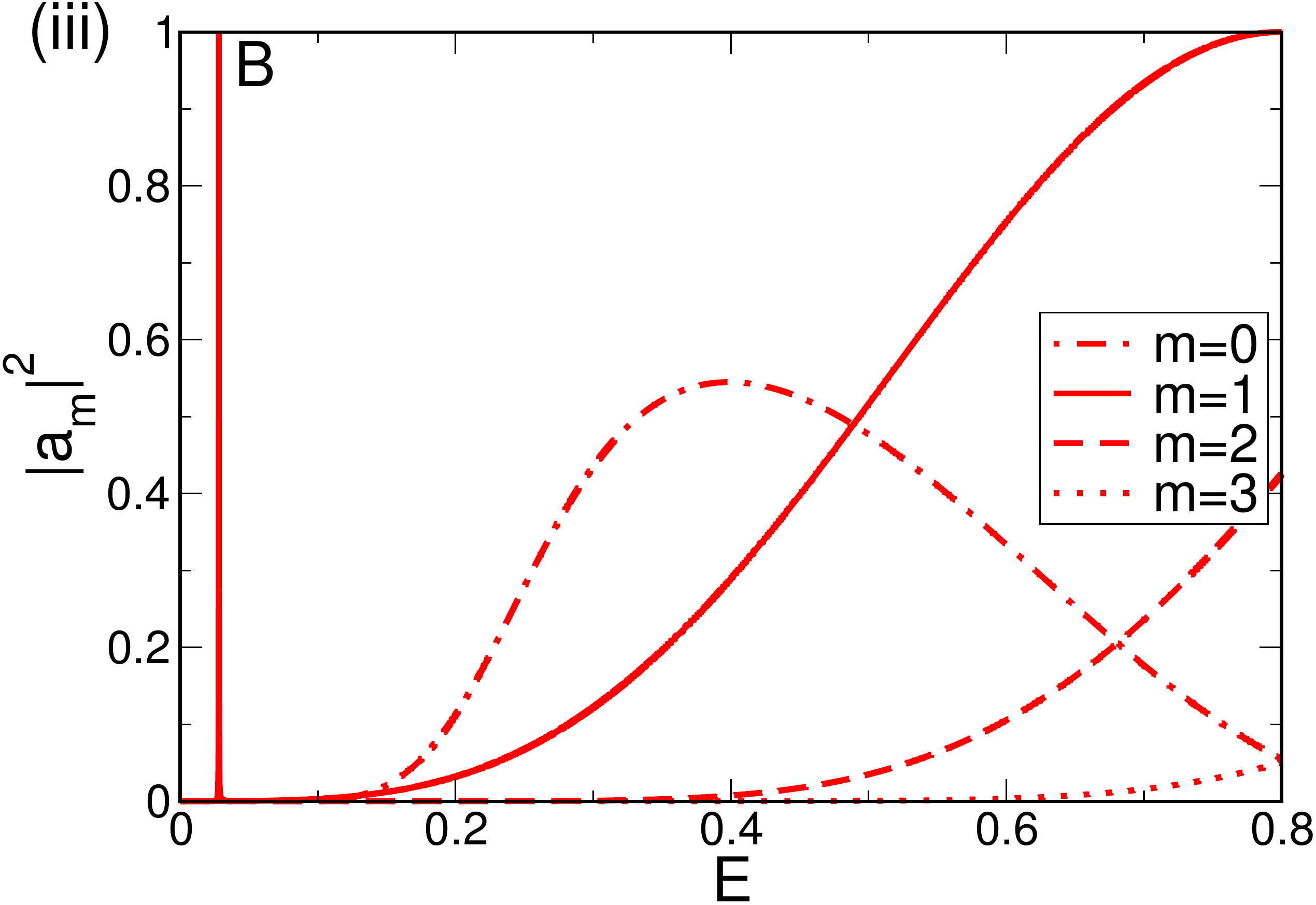}
\end{minipage}

\begin{minipage}{0.32\linewidth}
\includegraphics[width=1\linewidth]{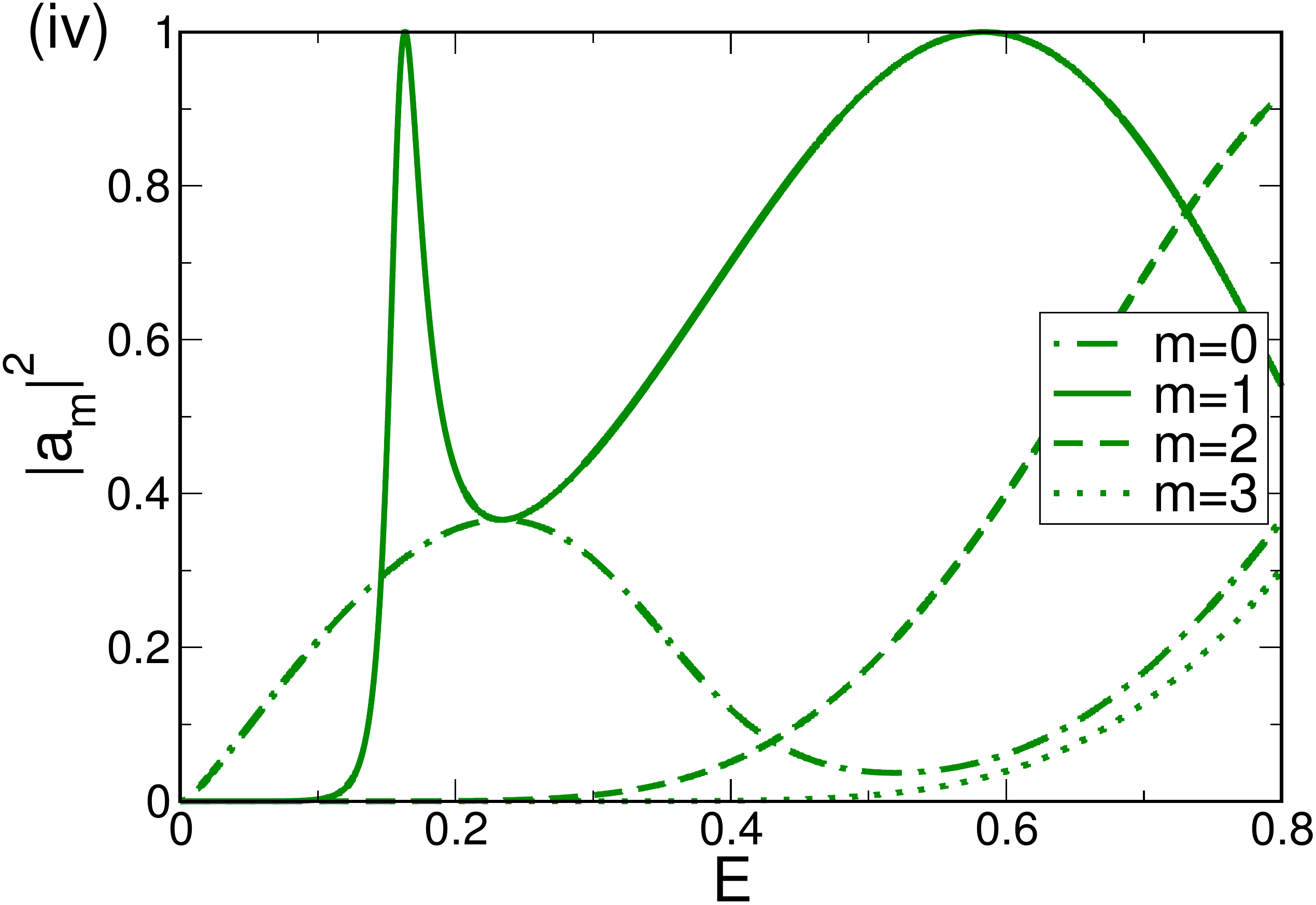}
\end{minipage}
\begin{minipage}{0.32\linewidth}
\includegraphics[width=1\linewidth]{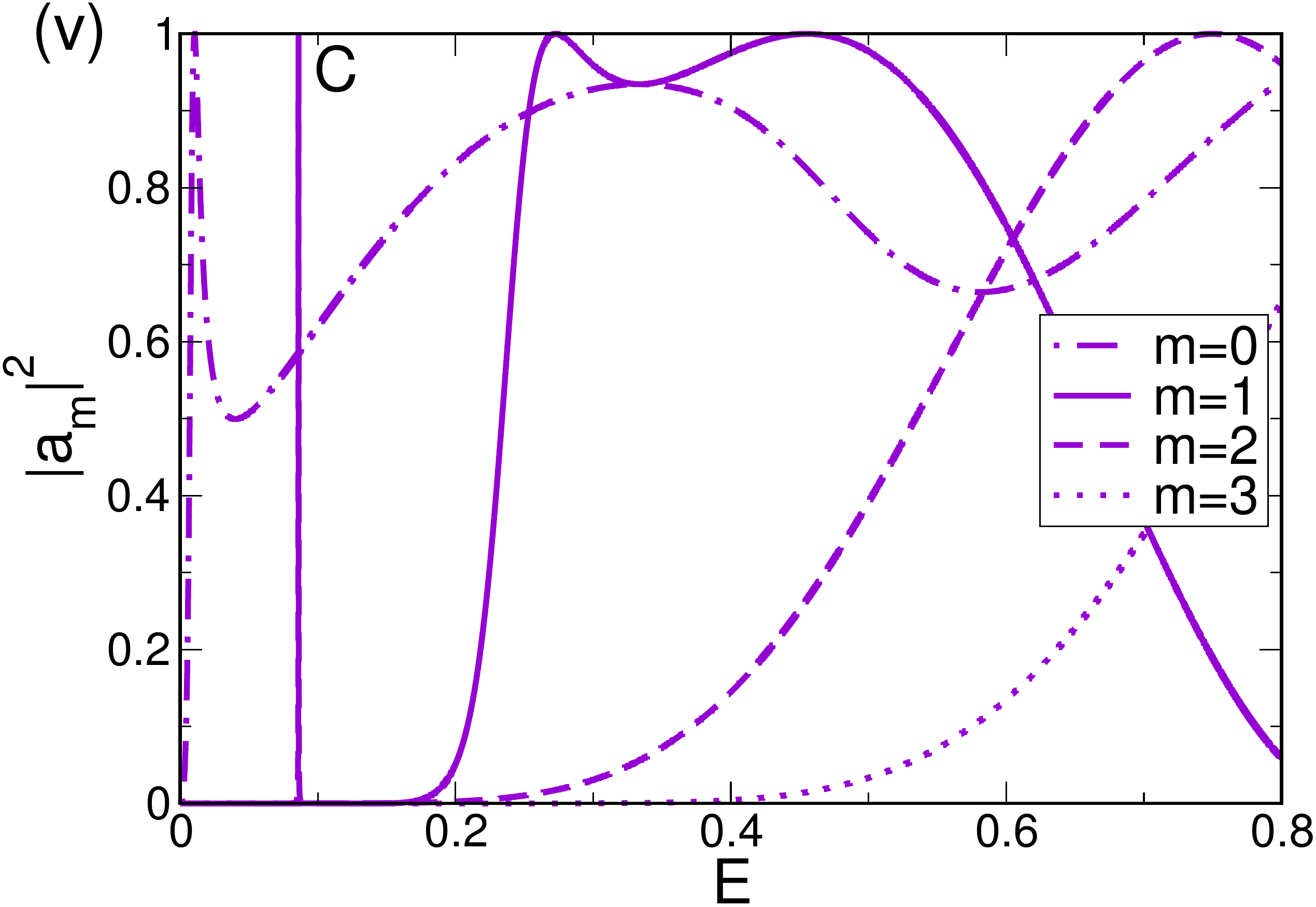}
\end{minipage}
\begin{minipage}{0.32\linewidth}
\includegraphics[width=1\linewidth]{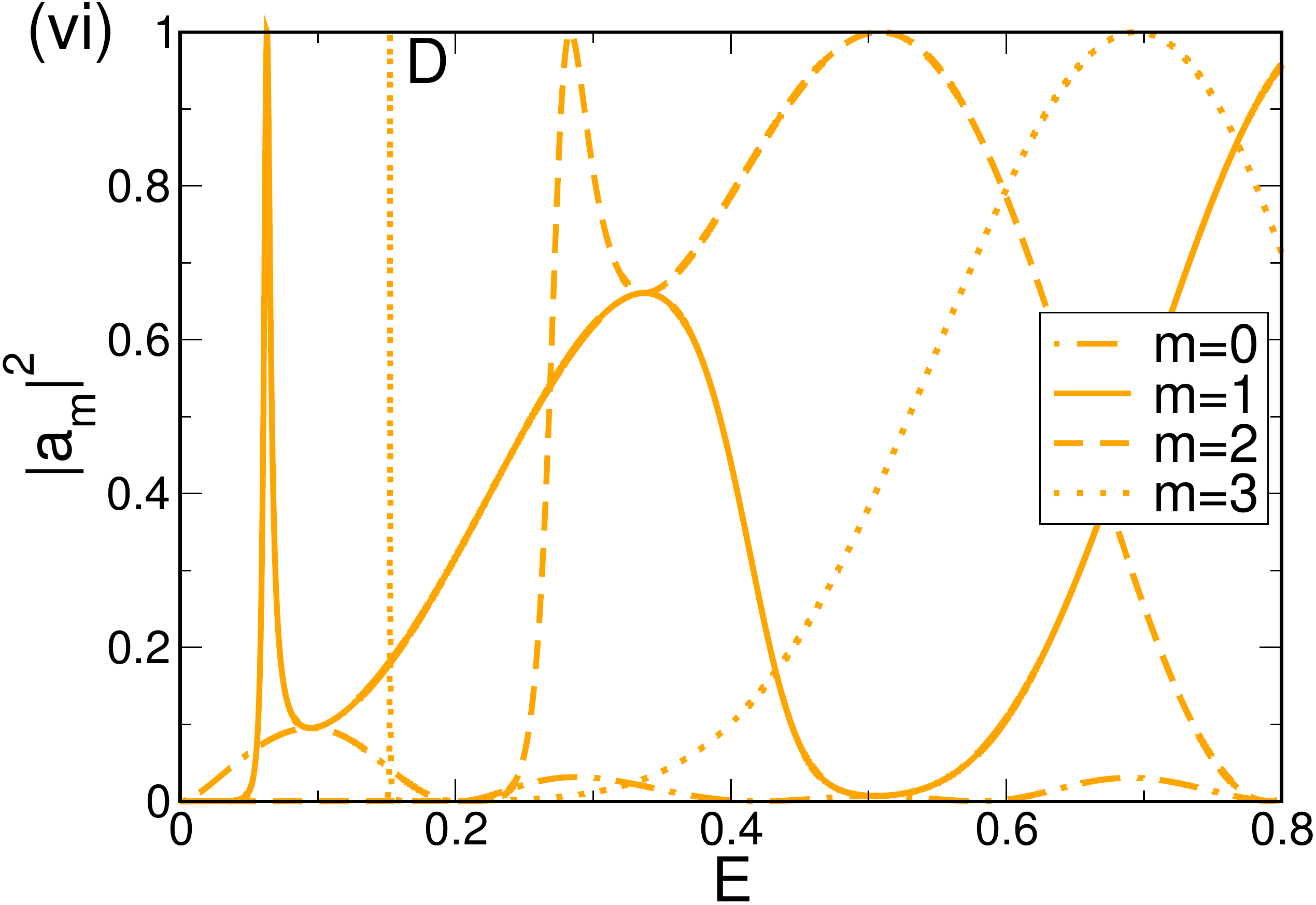}
\end{minipage}

\caption{Square modulus of the scattering coefficients $|a_m|^2$ for $m=0,1,2,3$ as a function of the energy $E$. The dot potential is $V=1$ and the radii are (i) $R=2$, (ii) $R=3$, (iii) $R=4$, (iv) $R=5$, (v) $R=5.75$, and (vi) $R=7.75$.}
\label{figure4}
\end{figure}

To further analyse the scattering for $E<V$ we show in Fig. \ref{figure3} the scattering efficiency as a function of the electron energy. We consider dots with small radii. In this regime sharp and broad maxima appear in $Q$ for specific $E$. They are due to the resonant excitation of normal modes of the dot. To identify the resonances we show in Fig. \ref{figure4} the energy dependence of the square modulus of the scattering coefficients $|a_m|^2$. Close to $E=0$ only the lowest scattering coefficient $a_0$ is non-zero. With increasing energy scattering coefficients of higher order contribute. For larger energy the $|a_m|^2$ tend to an oscillatory behaviour. For not too large $E$, however, the successive onset of modes is interspersed with sudden, sharp peaks of different $|a_m|^2$. These resonances of normal modes of the dot lead to the sharp peaks in $Q(E)$ found in Fig. \ref{figure3}.

\begin{figure}
\begin{minipage}{0.49\linewidth}
\includegraphics[width=1\linewidth]{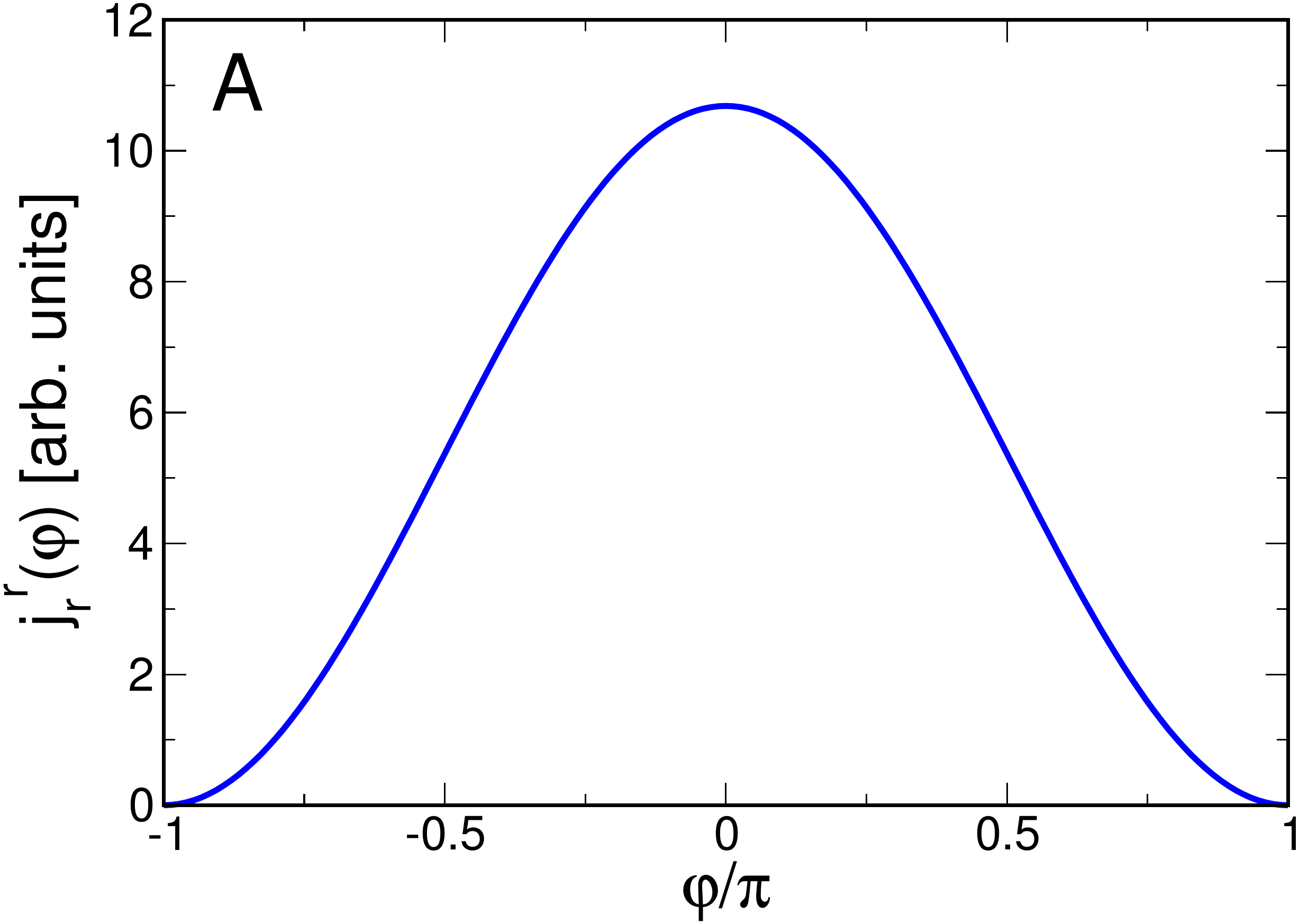}
\end{minipage}
\begin{minipage}{0.49\linewidth}
\includegraphics[width=1\linewidth]{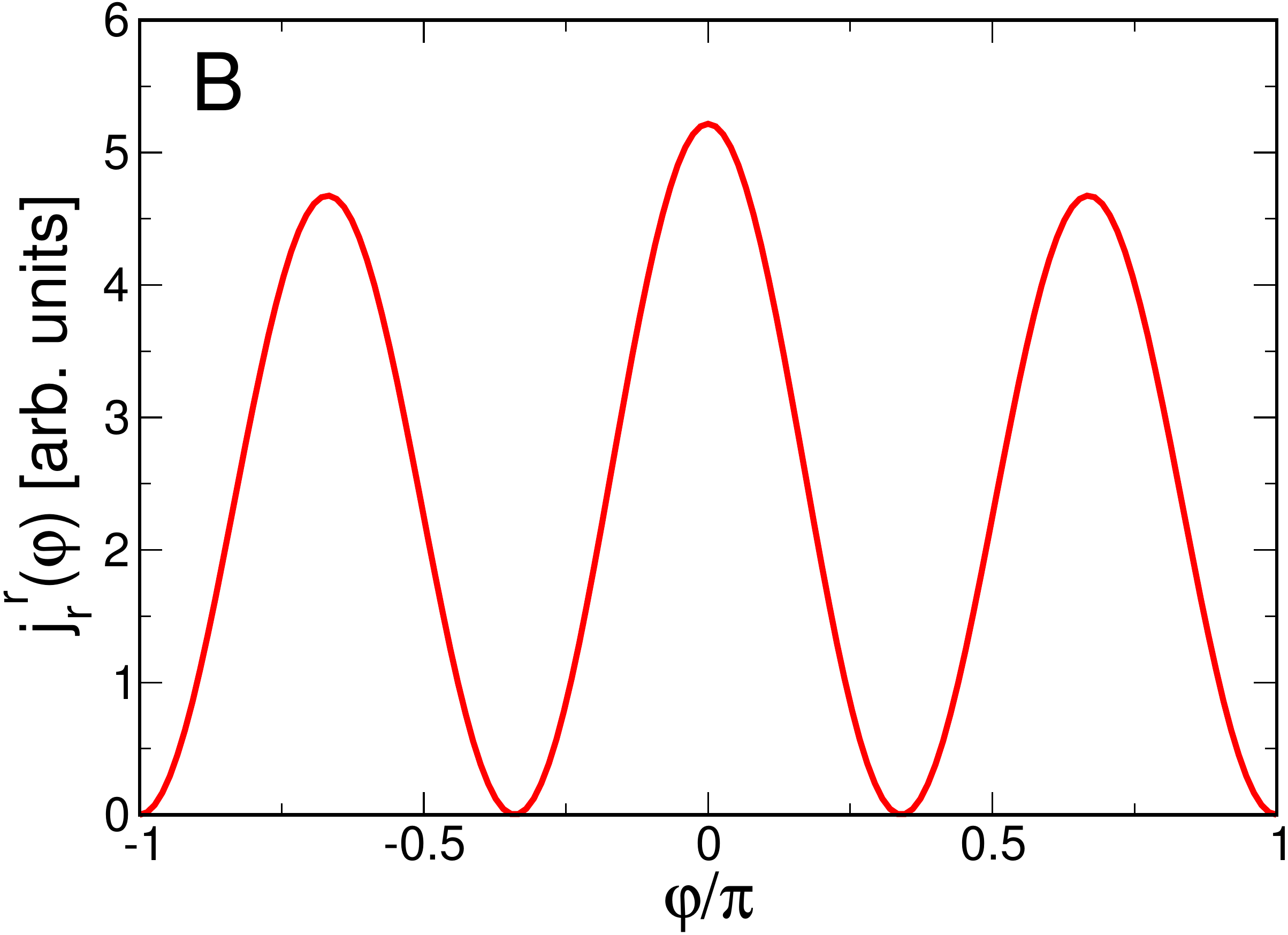}
\end{minipage}

\begin{minipage}{0.49\linewidth}
\includegraphics[width=1\linewidth]{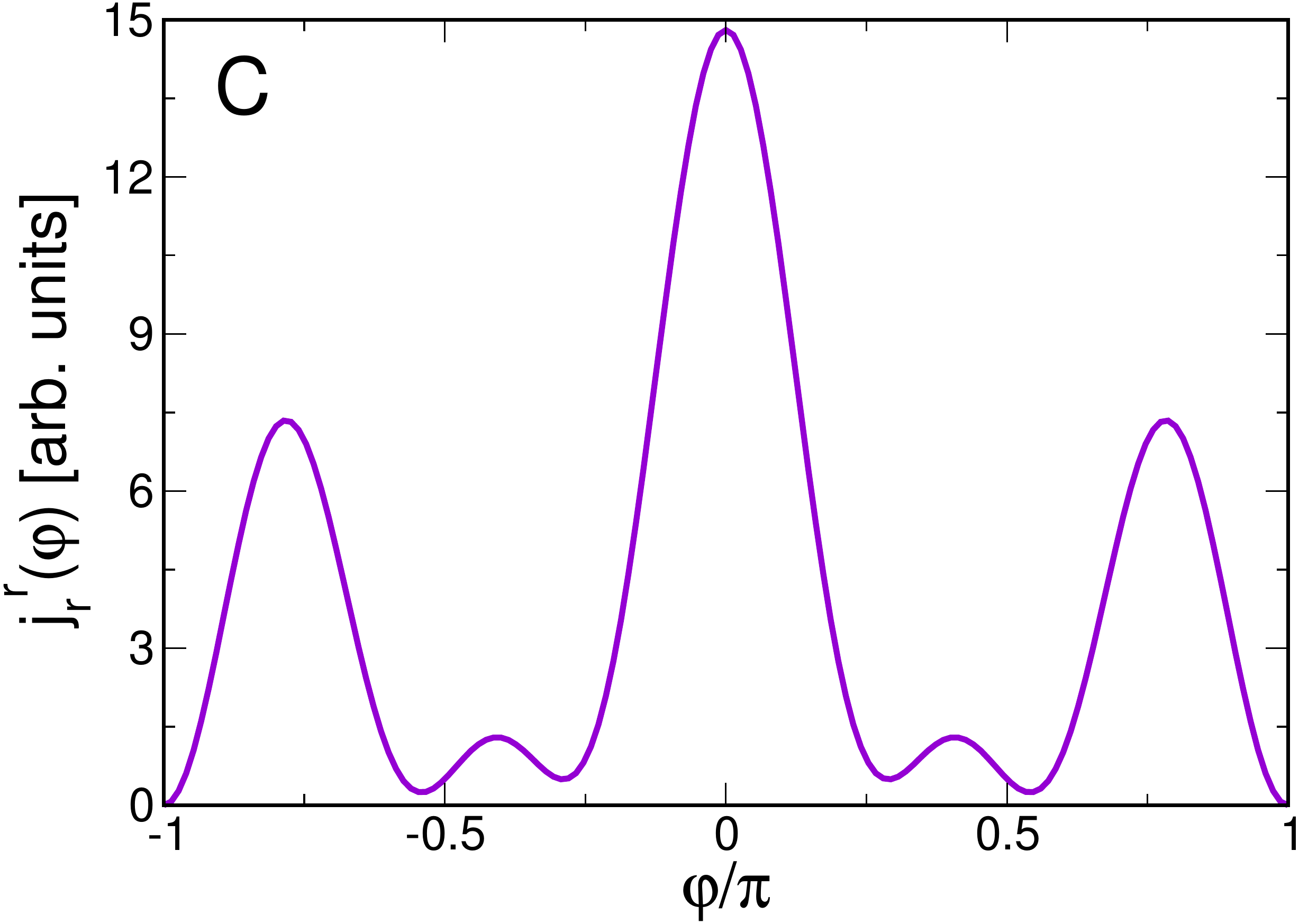}
\end{minipage}
\begin{minipage}{0.49\linewidth}
\includegraphics[width=1\linewidth]{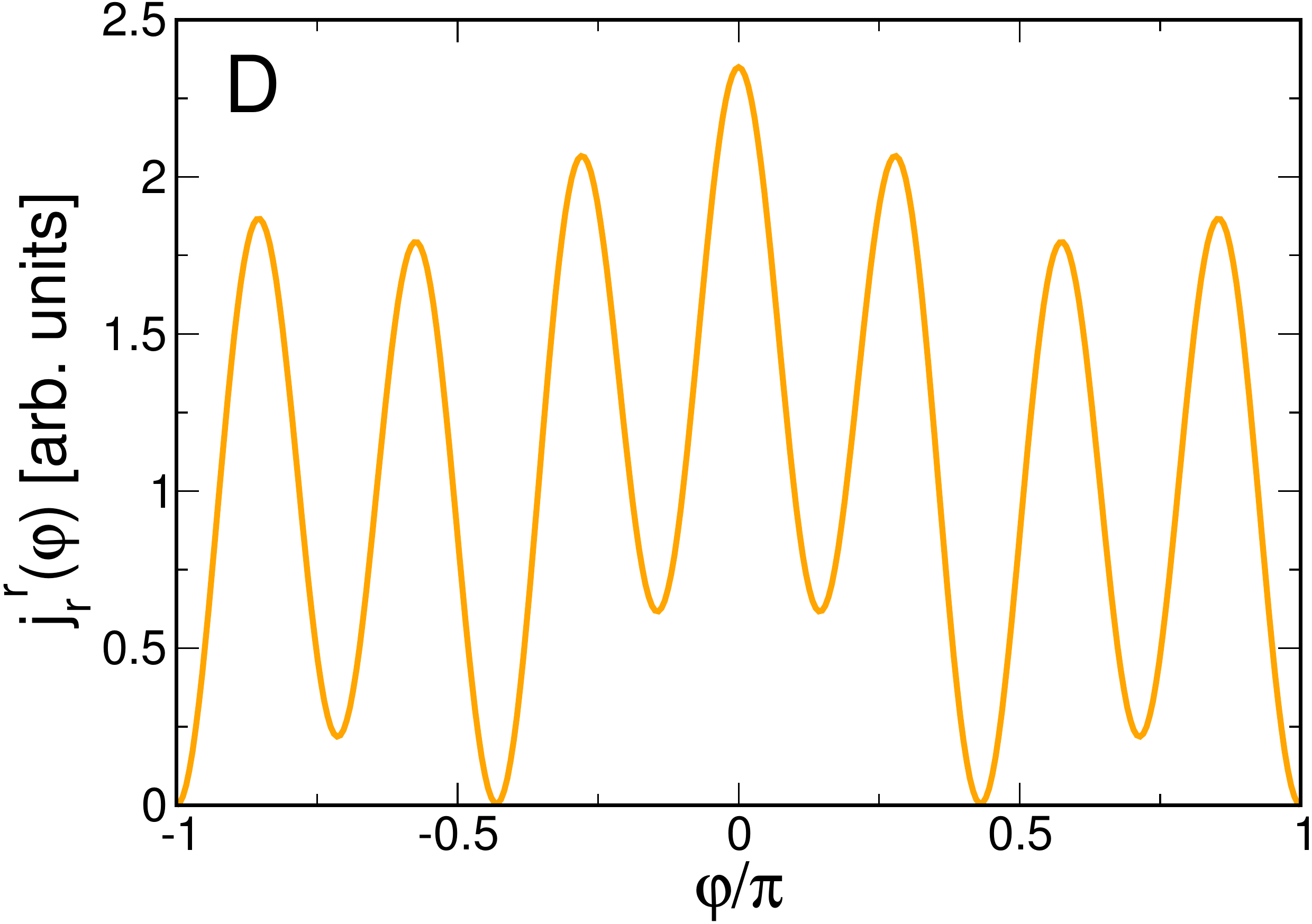}
\end{minipage}

\caption{Radial component of the far-field scattered current $j_r^r$ as a function of the angle $\varphi$ for (A) $R=3$ and $E=0.0704$, (B) $R=4$ and $E=0.0283$, (C) $R=5.75$ and $E=0.086$, and (D) $R=7.75$ and $E=0.1524$.}
\label{figure5}
\end{figure}

\begin{figure}
\begin{minipage}{0.49\linewidth}
\rotatebox{270}{\includegraphics[width=1\linewidth]{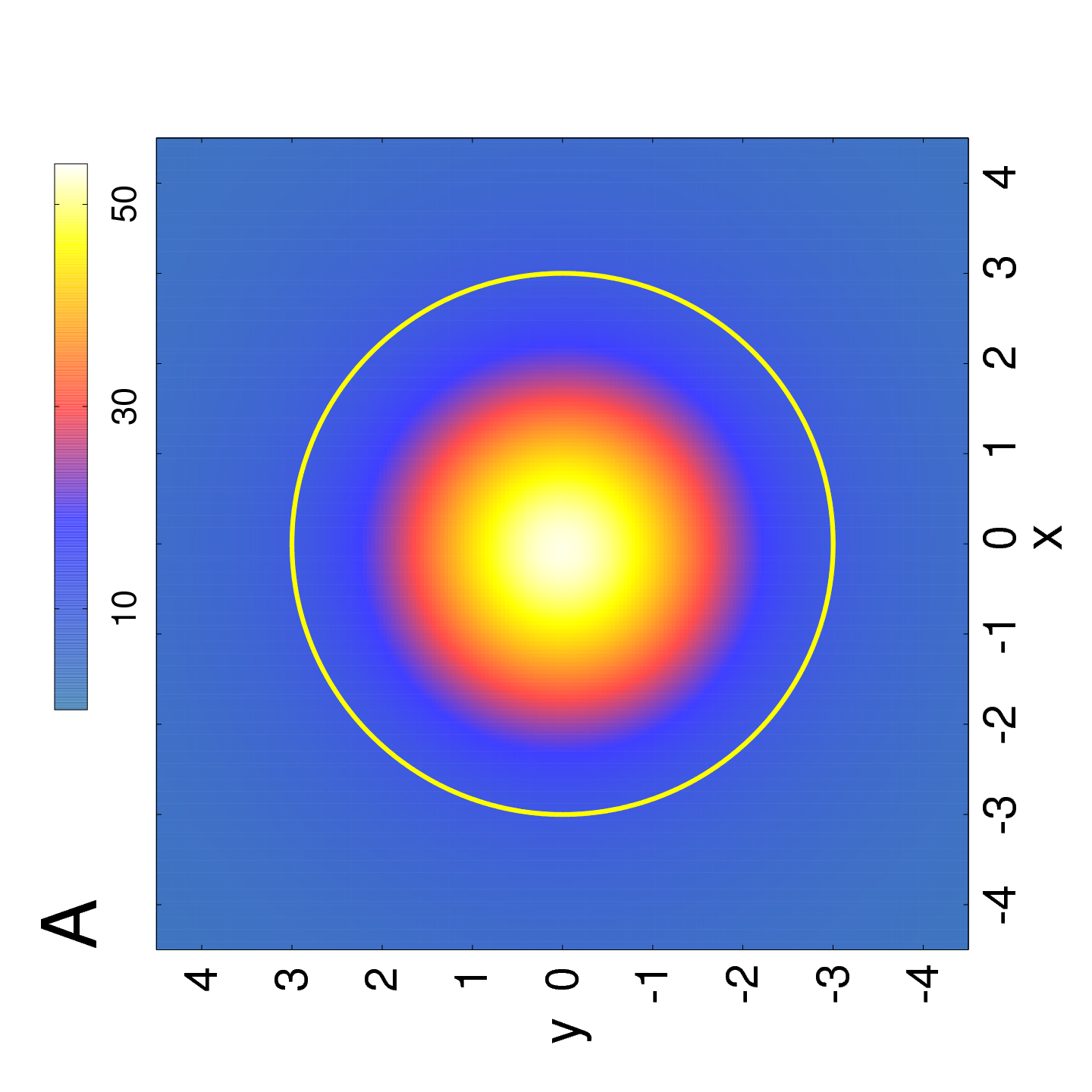}}
\end{minipage}
\begin{minipage}{0.49\linewidth}
\rotatebox{270}{\includegraphics[width=1\linewidth]{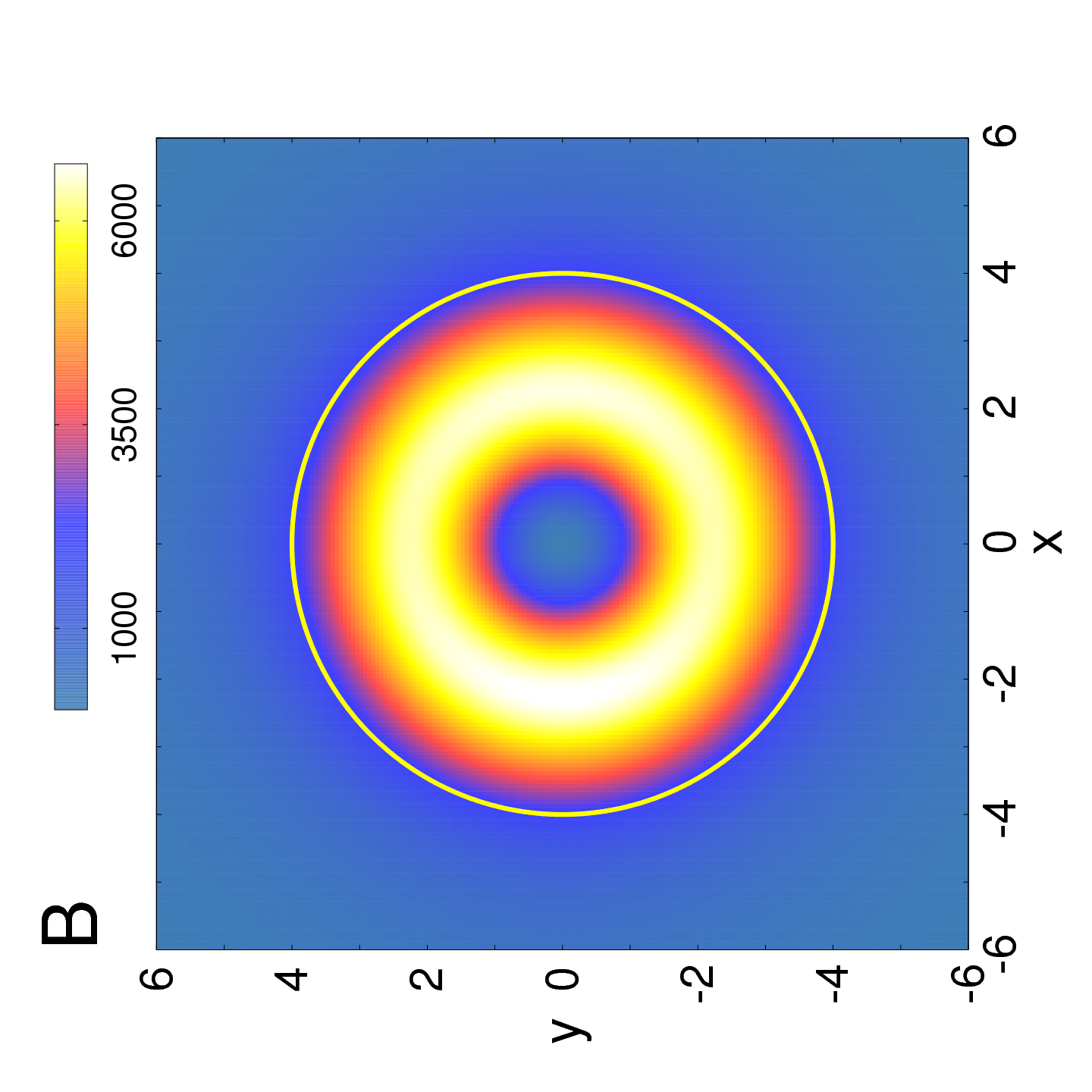}}
\end{minipage}

\begin{minipage}{0.49\linewidth}
\rotatebox{270}{\includegraphics[width=1\linewidth]{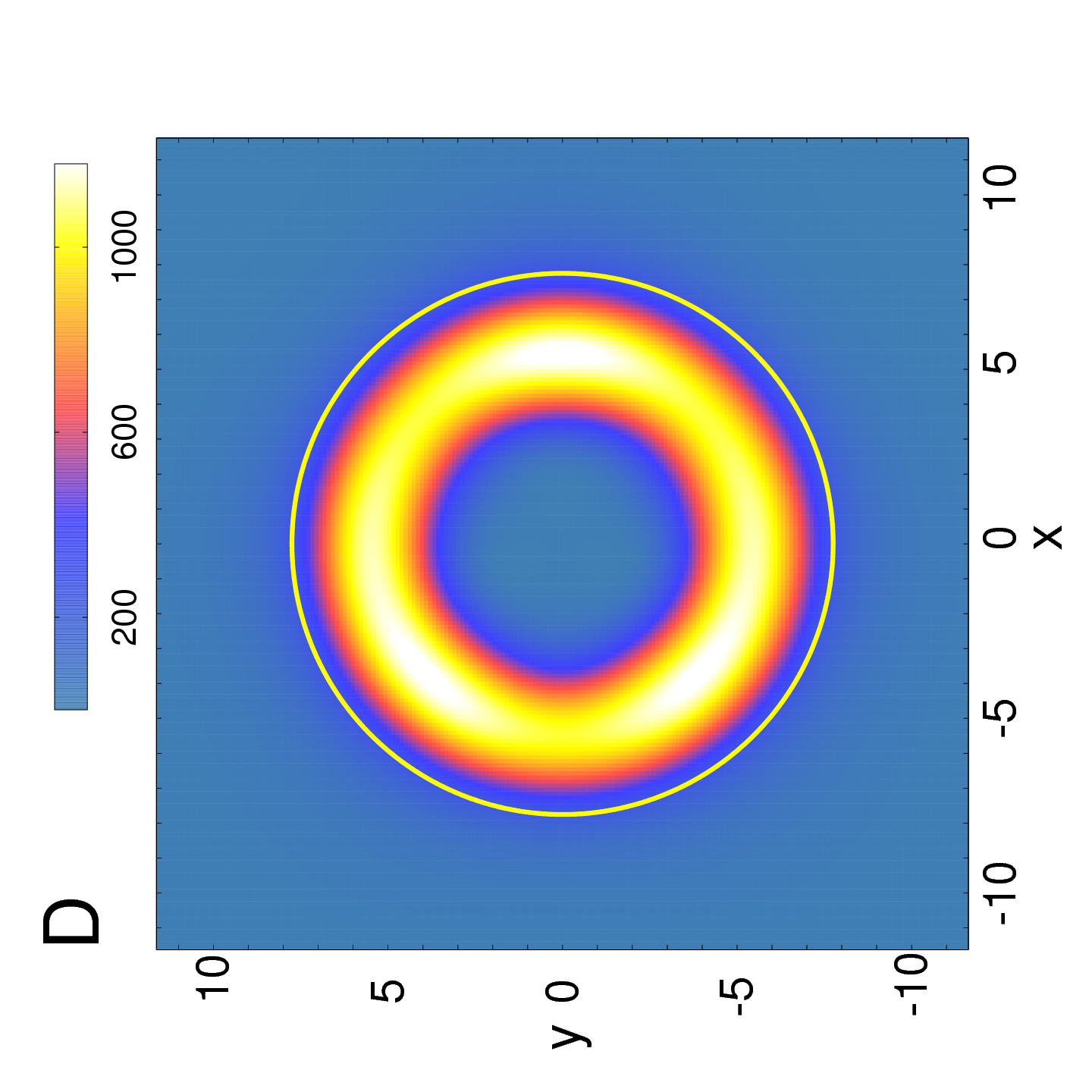}}
\end{minipage}
\begin{minipage}{0.49\linewidth}
\rotatebox{270}{\includegraphics[width=1\linewidth]{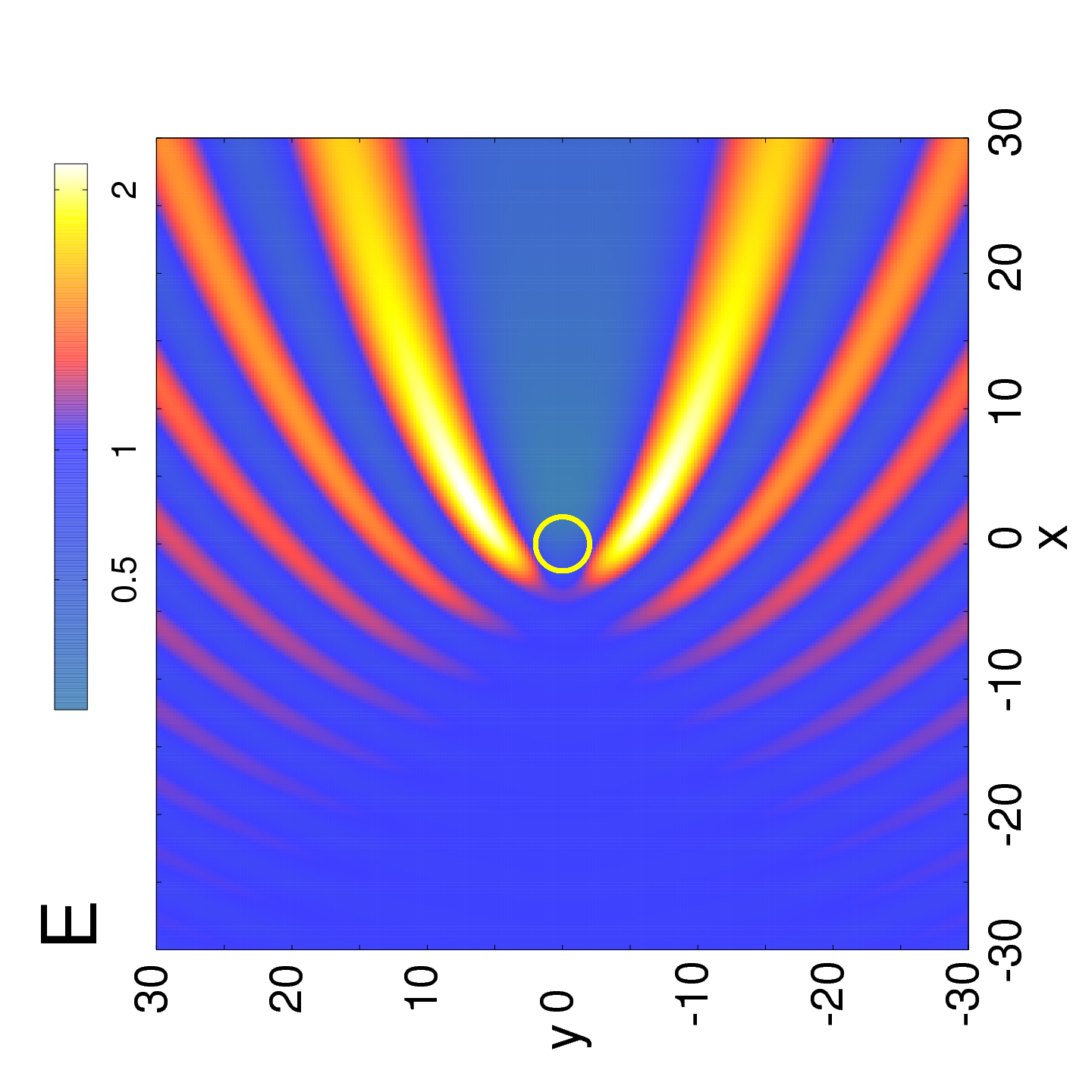}}
\end{minipage}

\caption{Spatial density profile $n=\psi^\dagger \psi$ in the vicinity of the dot for (A) $R=3$ and $E=0.0704$, (B) $R=4$ and $E=0.0283$, (D)  $R=7.75$ and $E=0.1524$ and (E) $R=2$ and $E=0.8$. In all panels $V=1$.}
\label{figure6}
\end{figure} 

The angular scattering characteristic is given by the radial component of the reflected current $j_r^\text{r}(\varphi)$ in the far field. If only one normal mode of the dot is excited $j_r^\text{r}(\varphi)$ takes the form
\begin{equation}
j_r^\text{r}(\varphi)\sim |a_m|^2 \left[ \cos((2m+1)\varphi)+1\right].
\end{equation}
The normal mode $a_m$ has $2m+1$ preferred scattering directions. Panel A and B of Fig. \ref{figure5} confirm that the mode $a_0$ scatters mainly in the forward direction while the mode $a_1$ has the three preferred scattering directions $\varphi=0,\pm 2\pi /3$. 
If apart from the dominant mode other modes contribute too, deviations from this scattering characteristic appear. This is exemplified in panels C and D of Fig. \ref{figure5}. Of the five preferred scattering directions of the mode $a_2$ two are suppressed (panel C) while for the mode $a_3$ the seven preferred scattering directions are still observable but with different amplitudes (panel D). Note that the absence of backscattering in graphene always implies $j_r^\text{r}(\varphi=\pm \pi)=0$.

Resonant scattering by only one of the normal modes is also reflected by the electron density profile in the vicinity of the dot. Inside the dot the density is given by
\begin{equation}
n=\psi_\text{t}^\dagger \psi_\text{t}=  |b_m|^2 \left[(J_m(qr))^2+(J_{m+1}(qr))^2 \right].
\end{equation}
Note that due to their common denominator, resonances of $a_m$ and of $b_m$ coincide.
The density is radially symmetric. For the mode $a_0$ the density maximum is in the centre of the dot and for the modes $a_{m>0}$ the density is ring shaped. Panel A, B and D of Fig. \ref{figure6} show the electron density for the modes $a_0$, $a_1$ and $a_3$. Note that in panel D the small contribution of other modes of the dot to scattering leads to the slight asymmetry of the ring shaped electron density. Most importantly, the electron density inside the dot is dramatically increased which is a sign of temporary particle trapping at the scattering resonances. Finally, panel E of Fig. \ref{figure6} shows the density profile for off-resonant scattering. In this case no electron trapping at the dot is observed but diffraction at the dot leads to an interference pattern in the passing wave.

\section{summary}

In this work we have studied the scattering of a plane Dirac electron wave on a circular, electrostatically defined quantum dot in graphene. Due to the conical energy dispersion in graphene, electrons occupy non-evanescent states inside the dot even when their energy is below the dot potential. In particular, for a low energy of the incident electron, scattering resonances due to the excitation of normal modes of the dot appear. They are characterised by distinct preferred scattering directions. At the scattering resonances the electron density in the dot is strongly increased which indicates temporary electron trapping.

\section*{acknowledgements}

This work was supported by the DFG Priority Programme 1459 Graphene. 


\end{document}